\documentstyle[preprint,prl,aps]{revtex}

\begin{document}

\bibliographystyle{prsty}

\tighten

\title{Information losses in continuous variable quantum teleportation}

\author{Holger F. Hofmann}
\address{CREST, Japan Science and Technology Corporation (JST)\\ 
Research Institute for Electronic Science, Hokkaido 
University\\
Kita-12 Nishi-6, Kita-ku, Sapporo 060-0018, Japan}

\author{Toshiki Ide, and Takayoshi Kobayashi}
\address{Department of Physics, Faculty of Science, University of Tokyo\\
7-3-1 Hongo, Bunkyo-ku, Tokyo113-0033, Japan}

\author{Akira Furusawa}
\address{Department of Applied Physics, University of Tokyo\\
7-3-1 Hongo, Bunkyo-ku, Tokyo113-8656, Japan}

\date{\today}

\maketitle

\begin{abstract}
It is shown that the information losses due to the limited fidelity of
continuous variable quantum teleportation are equivalent to the losses
induced by a beam splitter of appropriate reflectivity. 
\end{abstract}

%\pacs{PACS numbers:
%03.67.Hk  % Quantum communication
%42.50.-p  % Quantum Optics
%89.70.+c  % Information Science
%}
\vspace{0.5cm}

Quantum teleportation allows the transmission of an unknown quantum state
by combining the nonlocal quantum coherence of entangled states with the
transmission of classical information obtained in a joint measurement
of the unknown input state and one part of the entangled pair \cite{Ben93}. 
Ideally, the classical 
information transmitted contains no information about the input state and 
the output state is exactly identical with the input state. However, this 
ideal form of quantum teleportation requires maximal entanglement. 
In continuous 
variable quantum teleportation \cite{Vai94,Bra98,Fur98}, only non-maximal 
entanglement is available.
As a consequence, the output state is not perfectly identical to the input
state while the statistics of the joint measurement depend on the properties
of the input state \cite{Hof00}. The classical information channel then 
carries information on the input state which may be extracted, e.g. in order 
to eavesdrop on a quantum communication channel \cite{Ral00}. 
The relationship between the measurement information extracted 
and the change of the quantum state can be described in terms of a measurement
dependent transfer operator \cite{Hof00}. 
In the following, the transfer operator describing the 
continuous variable teleportation process is derived and the equivalence
with a feedback compensated beam splitter is established. The information
obtained in quantum teleportation can then be identified with the reflected
amplitude at the beam splitter while the output state of the teleportation 
corresponds to the transmitted beam, displaced by the feedback. The loss of
quantum information due to the limited fidelity of the teleportation process 
is thus shown to be equivalent to the loss of quanta at a beam splitter. 

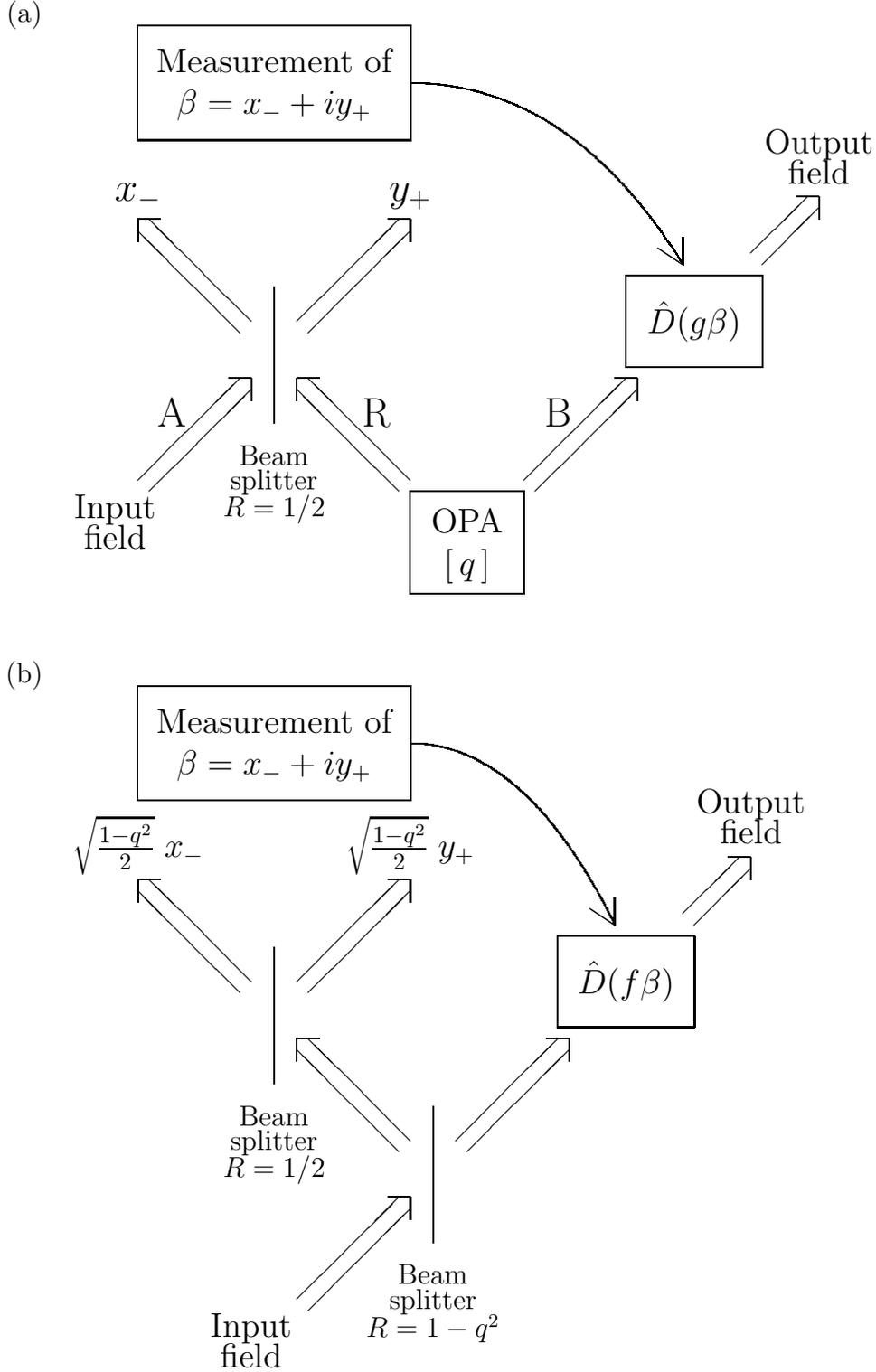
\begin{figure}

\setlength{\unitlength}{0.95pt}
\begin{picture}(400,320)
%\put(0,0){\framebox(400,300){}}
\put(10,300){\makebox(20,20){(a)}}

%\put(265,25){\line(-1,1){25}}
%\put(270,30){\line(-1,1){25}}
%\put(160,30){\line(1,1){25}}
%\put(165,25){\line(1,1){25}}
%\put(165,0){\makebox(100,30){Vacuum modes}}

\put(190,55){\framebox(50,45){}}
\put(190,77){\makebox(50,20){\large OPA}}
\put(190,57){\makebox(50,20){\large [$\, q \, $]}}

\put(190,105){\line(-1,1){45}}
\put(185,100){\line(-1,1){45}}
\put(140,150){\line(0,-1){10}}
\put(140,150){\line(1,0){10}}
\put(165,125){\makebox(20,20){\Large R}}

\put(240,105){\line(1,1){45}}
\put(245,100){\line(1,1){45}}
\put(290,150){\line(0,-1){10}}
\put(290,150){\line(-1,0){10}}
\put(245,125){\makebox(20,20){\Large B}}

\put(70,105){\line(1,1){45}}
\put(75,100){\line(1,1){45}}
\put(120,150){\line(0,-1){10}}
\put(120,150){\line(-1,0){10}}
\put(75,125){\makebox(20,20){\Large A}}
\put(40,86){\makebox(40,12){\large Input}}
\put(40,74){\makebox(40,12){\large field}}

\put(130,130){\line(0,1){60}}
\put(110,110){\makebox(40,12){Beam}} 
\put(110,98){\makebox(40,12){splitter}}
\put(110,86){\makebox(40,12){$R=1/2$}}

\put(120,175){\line(-1,1){45}}
\put(115,170){\line(-1,1){45}}
\put(70,220){\line(0,-1){10}}
\put(70,220){\line(1,0){10}}
\put(60,220){\makebox(20,20){\Large $x_-$}}

\put(140,175){\line(1,1){45}}
\put(145,170){\line(1,1){45}}
\put(190,220){\line(0,-1){10}}
\put(190,220){\line(-1,0){10}}
\put(180,220){\makebox(20,20){\Large $y_+$}}

\put(70,255){\framebox(120,50){}}
\put(80,275){\makebox(100,30){\large Measurement of}}
\put(80,255){\makebox(100,30){\large $\beta=x_-+i y_+$}}

\bezier{400}(190,280)(260,280)(310,200)
\put(310,200){\line(0,1){12}}
\put(310,200){\line(-3,2){11}}

\put(285,155){\framebox(60,40){\large $\hat{D}(g \beta)$}}

\put(340,205){\line(1,1){25}}
\put(345,200){\line(1,1){25}}
\put(370,230){\line(0,-1){10}}
\put(370,230){\line(-1,0){10}}

\put(350,247){\makebox(40,12){\large Output}}
\put(350,235){\makebox(40,12){\large field}}
\end{picture}

\vspace{-1cm}

\begin{picture}(400,320)
%\put(0,0){\framebox(400,300){}}
\put(10,300){\makebox(20,20){(b)}}

\put(200,60){\line(0,1){60}}
\put(180,40){\makebox(40,12){Beam}} 
\put(180,28){\makebox(40,12){splitter}}
\put(180,16){\makebox(40,12){$R=1-q^2$}}

\put(190,105){\line(-1,1){45}}
\put(185,100){\line(-1,1){45}}
\put(140,150){\line(0,-1){10}}
\put(140,150){\line(1,0){10}}

\put(210,105){\line(1,1){45}}
\put(215,100){\line(1,1){45}}
\put(260,150){\line(0,-1){10}}
\put(260,150){\line(-1,0){10}}

%\put(70,105){\line(1,1){45}}
%\put(75,100){\line(1,1){45}}
%\put(120,150){\line(0,-1){10}}
%\put(120,150){\line(-1,0){10}}

\put(140,35){\line(1,1){45}}
\put(145,30){\line(1,1){45}}
\put(190,80){\line(0,-1){10}}
\put(190,80){\line(-1,0){10}}

\put(100,16){\makebox(40,12){\large Input}}
\put(100,4){\makebox(40,12){\large field}}

\put(130,130){\line(0,1){60}}
\put(110,110){\makebox(40,12){Beam}} 
\put(110,98){\makebox(40,12){splitter}}
\put(110,86){\makebox(40,12){$R=1/2$}}

\put(120,175){\line(-1,1){45}}
\put(115,170){\line(-1,1){45}}
\put(70,220){\line(0,-1){10}}
\put(70,220){\line(1,0){10}}
\put(60,225){\makebox(20,20){\large $\sqrt{\frac{1-q^2}{2}}\; x_-$}}

\put(140,175){\line(1,1){45}}
\put(145,170){\line(1,1){45}}
\put(190,220){\line(0,-1){10}}
\put(190,220){\line(-1,0){10}}
\put(180,225){\makebox(20,20){\large $\sqrt{\frac{1-q^2}{2}}\; y_+$}}

\put(70,255){\framebox(120,50){}}
\put(80,275){\makebox(100,30){\large Measurement of}}
\put(80,255){\makebox(100,30){\large $\beta=x_-+i y_+$}}

\bezier{400}(190,280)(240,280)(280,200)
\put(280,200){\line(0,1){12}}
\put(280,200){\line(-3,2){11}}

\put(255,155){\framebox(60,40){\large $\hat{D}(f\beta)$}}

\put(310,205){\line(1,1){25}}
\put(315,200){\line(1,1){25}}
\put(340,230){\line(0,-1){10}}
\put(340,230){\line(-1,0){10}}

\put(320,247){\makebox(40,12){\large Output}}
\put(320,235){\makebox(40,12){\large field}}

\end{picture}

\vspace{0.5cm}

\caption{\label{setup} Comparison of the setups for continuous variable 
quantum teleportation (a) and the feedback compensated beam splitter (b).
$R$ denotes beam splitter reflectivities. All other parameters are as given 
in the text.}
\end{figure}

As illustrated in figure \ref{setup} (a),
quantum teleportation transfers an unknown quantum state in mode A
using an entangled state of a reference mode R and an output mode B.
For continuous variables, this is achieved by measuring the difference 
$\hat{x}_-=\hat{x}_A-\hat{x}_R$ and the sum 
$\hat{y}_+=\hat{y}_A+\hat{y}_R$ of the orthogonal quadrature components
of the input mode A and the reference mode R. The quantum state 
$\mid \psi_{\mbox{B}}(\beta)\rangle$ of the output mode B is then 
conditioned by the input state $\mid \psi_{\mbox{A}}\rangle$ in mode
A and the measurement result $\beta=x_- + i y_+$ which ideally defines an 
eigenstate $\mid \beta(A,R)\rangle$ of modes A and R. 
Reallistically, the
finite resolution due to limited detector efficiencies and other technical 
problems of the measurement will result in some additional noise,
which can be simulated by a classical Gaussian error. In the following, 
however, we will focus on the ideal quantum limit of the information transfer 
in order to determine the quantum state distortions originating from 
non-maximal entanglement. 
Using 
$\mid q(R,B)\rangle$ to denote the initial entangled state of modes 
R and B, this conditional state in B can be written as 
\begin{equation}
\mid \psi_{\mbox{B}}(\beta)\rangle = 
\langle \beta(A,R) \mid \psi_A\rangle \mid q(R,B)\rangle.
\end{equation}
Note that the output state $\mid \psi_{\mbox{B}}(\beta)\rangle$ is
not normalized, since the probability $P(\beta)$ of the measurement 
outcome $\beta$ is given by
\begin{equation}
P(\beta)=\langle \psi_{\mbox{B}}(\beta) \mid \psi_{\mbox{B}}(\beta) \rangle.
\end{equation}
Making use of the displacement 
operator $\hat{D}(\beta)$ and the photon number expansion of entanglement,
the eigenstates $\mid \beta(A,R)\rangle$ and the entangled state
$\mid q(R,B)\rangle$ can be expressed as
\begin{eqnarray}
\mid \beta(A,R)\rangle &=& \frac{1}{\sqrt{\pi}}
\sum_{n=0}^{\infty} \hat{D}_A(\beta) \mid n ; n \rangle_{A,R}
\nonumber \\ 
\mid q(R,B)\rangle &=& \sqrt{1-q^2}
\sum_{n=0}^{\infty} q^n \mid n ; n \rangle_{R,B},
\end{eqnarray}
where the entanglement coefficient $q$ provides a quantitative measure 
characterizing the degree of entanglement obtained by parametric 
amplification.
The quantum state $\mid \psi_{\mbox{B}}(\beta)\rangle$ of the output mode
B conditioned by the measurement of modes A and R is then given by
\begin{equation}
\mid \psi_B (\beta)\rangle = \sqrt{\frac{1-q^2}{\pi}} \sum_{n=0}^{\infty}
q^{n} \mid n\rangle \langle n \mid \hat{D}_A(-\beta) \mid \psi_A \rangle.
\end{equation} 
For $q=1$ this state is a copy of the input state displaced by a field
difference of $-\beta$. Therefore the final step of quantum teleportation is 
the reversal of this displacement by the addition of a coherent field 
amplitude $g \beta$ to obtain the final output state, 
$\mid \psi_{\mbox{out}}(\beta)\rangle = 
\hat{D}(g \beta) \mid \psi_B(\beta)\rangle $. The gain factor
$g$ allows an adjustment of the teleported amplitude \cite{Fur98}. $g=1$
reproduces the average input amplitude in the output and thus optimizes
the fidelity for the teleportation of high amplitude coherent states.
As discussed in a previous paper \cite{Hof00}, the conditional output 
of quantum teleportation can be described using the transfer operator 
$\hat{T}_g(\beta)$,
\begin{eqnarray}
\mid \psi_{\mbox{out}} (\beta)\rangle \hspace{0.2cm} &=& \hspace{1.0cm}  
\hat{T}_g(\beta) \mid \psi_A \rangle
\nonumber \\[0.3cm]
P(\beta)\hspace{0.7cm}&=& \hspace{0.2cm} \langle \psi_A \mid \hat{T}_g^\dagger(\beta)
\hat{T}_g(\beta) \mid \psi_A \rangle
\nonumber \\[0.2cm]
\mbox{with} &&
\hat{T}_g(\beta) = \sqrt{\frac{1-q^2}{\pi}} \sum_{n=0}^{\infty}
q^{n} \hat{D}(g \beta) \mid n\rangle \langle n \mid \hat{D}(-\beta).
\end{eqnarray}
This transfer operator characterizes the teleportation process of an
arbitrary quantum state by correlating the extracted information $\beta$
with the quantum information in the output state 
$\mid \psi_{\mbox{out}} (\beta)\rangle $. When applied to a coherent state 
$\mid \alpha \rangle$, the result reads
\begin{eqnarray}
\label{eq:alpha}
\hat{T}_g(\beta)\mid \alpha\rangle &=& \sqrt{\frac{1-q^2}{\pi}} 
\exp\left(-(1-q^2)\frac{|\alpha-\beta|^2}{2}\right) \nonumber \\ &&
\times 
\exp\left((1-g q)\frac{\alpha\beta^*-\beta\alpha^*}{2}\right)  
\mid q \alpha +(g-q) \beta \rangle.
\end{eqnarray}
For coherent states, the output is also a coherent state with an
amplitude give by the sum of the attenuated input amplitude 
$q \alpha$ and a measurement dependent displacement of $(g-q)\beta$. 
Since the quantum state in the output depends on the randomly varying 
measurement result $\beta$, the teleportation output is generally a 
statistical mixture of different coherent states. Only in the 
special case of $g=q$, the amplitude $q\alpha$ of the coherent output
state does not depend on $\beta$ and the output is a well defined pure 
state even if the output is averaged over all measurement results $\beta$.

The attenuation of the signal amplitude described by equation
(\ref{eq:alpha}) corresponds to the losses induced by a beam splitter
with a reflectivity of $1-q^2$. 
For the special case of $g=q$, this 
property of teleportation has been pointed out previously by Polkinghorne
and Ralph \cite{Pol99,Ral99}, based on an analysis of the 
quantum fluctuations in the teleportation. In the following, we will 
generalize this analogy by deriving the proper transformation of quantum 
states in a beam splitter measurement and considering the possibility 
of compensating the beam splitter losses by feedback. This formalism 
includes all the details necessary for an evaluation of the information 
obtained on the system and the minimal back-action on the signal field.
If the reflected amplitude of 
$\sqrt{1-q^2}\; \alpha$ is measured by eight port homodyne detection, the 
corresponding positive operator valued measure is given by projections
onto the non-normalized, non-orthogonal coherent states
\begin{eqnarray}
\mid P(\beta)\rangle &=& \sqrt{\frac{1-q^2}{\pi}}
\; \mid\, \sqrt{1-q^2} \beta \;\rangle
\nonumber \\ \mbox{with} && 
\int d^2\beta \mid P(\beta)\rangle\langle P(\beta)\mid=\hat{1},
\end{eqnarray}
and the transmitted state reads
\begin{eqnarray}
\mid \psi_{\mbox{trans}}(\beta)\rangle &=& \sqrt{\frac{1-q^2}{\pi}}\;
\langle \sqrt{1-q^2} \beta \mid  \sqrt{1-q^2} \alpha \rangle 
\; \mid q \alpha \rangle
\nonumber \\\ &=& \sqrt{\frac{1-q^2}{\pi}} 
\exp\left(-(1-q^2)\frac{|\alpha-\beta|^2}{2}\right) \nonumber \\ &&
\times 
\exp\left((1-q^2)\frac{\alpha\beta^*-\beta\alpha^*}{2}\right)  
\mid q \alpha \rangle,
\end{eqnarray}
where only the probability amplitude of 
$\mid \psi_{\mbox{trans}}(\beta)\rangle$ depends on the outcome $\beta$.
This result corresponds to equation (\ref{eq:alpha}) if the gain $g$ is
equal to the entanglement coefficient $q$. The effects of quantum teleportation
at a gain of $g=q$ are therefore identical to the effects of a field 
measurement by eight port homodyne detection performed on the reflected 
part of the signal field using a beam splitter of reflectivity $R=1-q^2$. 
At other gain coefficients, quantum teleportation is equivalent to a 
feedback compensated beam splitter measurement in which the losses 
induced in the transmitted beam are compensated using a linear feedback 
based on the measurement 
result $\beta$ obtained from the reflected light \cite{note}. 
At a feedback amplitude
of $f \beta$, the output state of the feedback compensated beam splitter
reads
\begin{equation}
\label{eq:equal}
\hat{D}\left( f \beta \right) \mid \psi_{\mbox{trans}}(\beta)\rangle = 
\hat{T}_{g=f+q}(\beta) \mid \alpha \rangle.
\end{equation}
In particular, a gain of
$g=1$ corresponds to a beam splitter feedback amplitude of $(1-q)\beta$.
In this case, the measurement operator of the beam splitter setup is
hermitian, minimizing the back-action of the measurement to the minimal 
noise required by the Heisenberg principle \cite{Wis95}. 

Since all quantum states may be expanded in terms of coherent states,
equation (\ref{eq:equal}) proofs the equivalence of continuous variable
quantum teleportation and feedback compensated beam splitting with respect to
both the changes in the transmitted state and the information obtained in the
measurement of $\beta$. 
In the special case of $g=q$, no additional photons
are created in the teleportation process, indicating that all photons
emitted into the output field B by the parametric amplifier are reabsorbed
in the displacement transformation. This effect allows a teleportation of the 
vacuum with a fidelity of one, making a more reliable distinction of 
signal pulses from a vacuum background possible. The transmission probability
for photons teleported at $g=q$ is equal to $q^2$. The loss of quantum 
information in continuous variable teleportation can thus be expressed in
terms of photon losses. In an experimental realization of quantum 
teleportation, the case of $g=q$ can be used to characterize the performance
of the setup. Specifically, the point at which $g=q$ can be found by 
minimalizing the output intensity at a vacuum input. The remaining intensity 
at that point arises from 
the finite resolution of the measurement, 
imperfect phase matching, and similar technical problems in the 
optical setup. It is then possible to separate quantum noise effects from
the classical noise contributions in the teleportation setup.

At $g>q$, the loss of quantum information is compensated 
by the classical information obtained from the measurement of $\beta$. 
However, the original quantum state cannot be restored by this purely 
classical manipulation, limiting the achievable fidelity to a value well 
below one. At $g<q$, the displacement actually reduces the quantum information
in the output further, until at $g=0$, there is no correlation between the 
input state and the output density matrix formed by integrating over all 
measurement results $\beta$. In general, 
the effect of quantum teleportation on the one photon state 
$\mid 1 \rangle$ is given by
\begin{equation}
\hat{T}_g(\beta)\mid 1 \rangle = \sqrt{\frac{1-q^2}{\pi}} \exp\left(-(1-q^2)
\frac{|\beta|^2}{2}\right) \hat{D}\left((g-q)\beta\right)
\left((1-q^2)\beta^*\mid 0 \rangle + q \mid 1 \rangle\right). 
\end{equation}
This representation of the teleported state has only two components, 
corresponding to a displaced vacuum and a displaced photon number state,
respectively. At $g=q$, the displacement is zero and the two components 
correspond to the actual loss or transmission of the photon. At other
gain factors, the coherent displacement can generate photon numbers $n>1$
in the output.  
Figure \ref{setup} shows a schematic comparison of the experimental setups 
for quantum teleportation setup and for the compensated beam splitter. 
Both methods employ linear transformations
on the input field mode and two vacuum modes, extracting information on
the unknown input field from the homodyne detection measurements on two
of the output modes. However, in quantum teleportation, the only physical
connection between the input field and the output field is given by the
measurement dependent displacement. While the beam splitter transmits 
the attenuated input field by a direct physical interaction, quantum 
teleportation achieves the same result by combining the entanglement with 
the classical information $\beta$. The entanglement coefficient $q$ is 
the measure of non-maximal entanglement which corresponds to the
attenuation of the transmitted amplitude at an equivalent beam splitter 
\cite{Rei89}. 

In conclusion, the analysis of the transfer operator $\hat{T}_g(\beta)$
shows that the information transfer in quantum teleportation is essentially 
equivalent to a feedback compensated beam splitter. This result clarifies
the nature of information losses in quantum teleportation and allows an
assessment of the information extracted with respect to applications such
as continuous variable eavesdropping \cite{Ral00}. Moreover, the analogy 
provides a quantification of the information transfer properties of 
non-maximal entanglement in terms of the photon transmission probability
$q^2$ 
and simplifies the derivation of quantum coherent transfer properties 
for few photon inputs, e.g. for the entanglement swapping scheme discussed
in \cite{Pol99}.

One of us (HFH) would like to acknowledge support from the Japanese 
Society for the Promotion of Science, JSPS.

%=========================================================

%=========================================================

\end{document}